\documentclass[aps,pra,reprint,twocolumn,superscriptaddress,amsmath,amssymb,citeautoscript]{revtex4-1}
\usepackage{amsmath}
\usepackage{amssymb}
\usepackage{graphicx}
\usepackage{color}
\usepackage{bm}
\usepackage{float}

\usepackage{epstopdf}
\usepackage{grffile}
\DeclareGraphicsExtensions{.eps}
\renewcommand{\vec}[1]{\bm{#1}}

\newcommand{\e}{{\rm e}}
\newcommand{\ii}{{\rm i}}

\usepackage{xspace}

\newcommand{\VL}{{VL$_{1/2}$}\xspace}
\newcommand{\Mz}{{M$_{0}$}\xspace}
\newcommand{\Mzp}{{M$_{0}'$}\xspace}
\newcommand{\Mpi}{{M$_{\pi/2}$}\xspace}

\newcommand{\jc}{{j_c}}

\begin{document}

\title{Quantum phases of strongly-interacting bosons on a two-leg Haldane ladder}
\author{S. Greschner}
\affiliation{Institut f\"ur Theoretische Physik, Leibniz Universit\"at Hannover, 30167~Hannover, Germany} 
\affiliation{Department of Quantum Matter Physics, University of Geneva, 1211 Geneva, Switzerland} 
\author{F. Heidrich-Meisner}
\affiliation{Department of Physics and Arnold Sommerfeld Center for Theoretical Physics, Ludwig-Maximilians-Universit\"at M\"unchen, 80333 M\"unchen, Germany}

\begin{abstract}
We study the ground-state physics of a single-component Haldane model on a hexagonal two-leg ladder geometry with a  particular focus on 
strongly interacting bosonic particles. We concentrate  our analysis on the regime of less than one particle per unit-cell. As a main result,  we observe several Meissner-like and vortex-fluid phases both for a superfluid as well as a Mott-insulating background. Furthermore, we show that for strongly interacting bosonic particles an unconventional vortex-lattice phase emerges, which is stable even in the regime of hardcore bosons. We discuss the mechanism for its stabilization for finite interactions by a means of an analytical  approximation.
We show how the different phases may be discerned by measuring the nearest- and next-nearest-neighbor chiral currents as well as their characteristic momentum distributions.
\end{abstract}
\date{\today}

\maketitle
\section{Introduction}

With the rapid progress in the realization of synthetic magnetism in ultracold atomic gases during  recent years, experiments in this field are now at the cusp of complementing the theoretical approaches and solid-state experiments on topological effects in {\it strongly correlated} quantum systems~\cite{Galitski2013, Goldman2016}.
At the same time, seminal advances in experiments with irradiated graphene~\cite{oka2009,wang2013} or photonic lattices~\cite{Hafezi2011, Rechtsman2013, Mittal2016} have shown the availability of these technologies for the investigation of topological states of matter as well.
So far, experiments have succeeded with several proof of concept measurements of various topological effects, most of which, however, studied  noninteracting particles. Among these  efforts, we mention the quantum engineering of various Hofstadter-Harper like models~\cite{Hof1976} with staggered~\cite{Aidelsburger2011, Struck2012} or rectified fluxes~\cite{Miyake2013, Aidelsburger2013} in ultracold atoms. 
 Highly non-trivial properties can be measured in these experiments  such as chiral currents~\cite{Atala2014, Stuhl2015, Mancini2015}, Chern numbers~\cite{Aidelsburger2015, Sugawa2016, Lohse2017} or Berry curvatures~\cite{Atala2013,Li2016, Flaschner2016}.
The theoretical understanding of interaction effects of such models, however, remains challenging and has triggered numerous studies in this field. For Hofstadter-Harper like models those include, e.g., predictions of interacting (fractional) Chern insulators \cite{Sorensen2005, Moeller2009,Regnault2011,Neupert2011,Kjaell2012,Moeller2015} and other unconventional quantum states \cite{Lim2008, Orth2012,Radic2012,Cole2012,Zhou2013}.

\begin{figure}[b]
\centering
\includegraphics[width=0.7\linewidth]{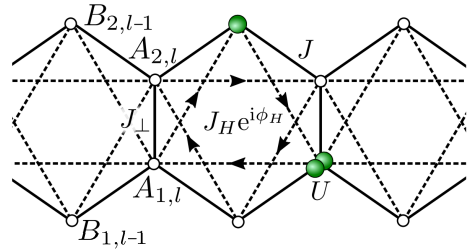}
\caption{(Color online) Sketch of the Haldane ladder and the model Eq.~\eqref{eq:HH} for interacting bosonic atoms. Throughout the  paper we set $J_\perp=J$. The
strength of the next-nearest neighbor tunneling matrix elements is $J_H$ (dashed lines), the phase attached to these links is $\phi_{H}$. The onsite interaction strength is denoted
by $U$. The unit cell contains four sites, denoted by $A_{1,\ell}$, $A_{2,\ell}$, $B_{1,\ell}$, $B_{2,\ell}$, where the first index labels the upper(lower) leg of the ladder. 
}
\label{fig:sketch}
\end{figure}

Another paradigmatic example of a model with nontrivial topological  phases is the famous Haldane model~\cite{Haldane1988}, given by the Hamiltonian
\begin{align}
H_H &= J \sum_{\langle \ell,\ell' \rangle} (c^\dagger_{\ell} c_{\ell'} +{\rm H.c.})\nonumber\\ 
&+J_H \sum_{\langle\langle \ell ,\ell' \rangle\rangle} (\e^{\ii\phi_H} c^\dagger_{\ell} c_{\ell'} +{\rm H.c.})\,,
\label{eq:HH}
\end{align}
where $c_{\ell}$ ($c_{\ell}^\dagger$) describes a single-component fermionic or bosonic annihilation (creation) operator, with 
$\langle \ell,\ell' \rangle$ denoting nearest neighbors and $\langle\langle \ell,\ell' \rangle \rangle$ next-nearest neighbors. A sketch of the model is shown in Fig.~\ref{fig:sketch}.
Contrary to the example of topological states of matter realized in an electronic system with a strong magnetic field~\cite{Hof1976}, here, no net flux pierces the unit-cell of the lattice and, hence, translational symmetry is  not   explicitly broken.
In spite of its apparent complexity - the need of complex next-nearest neighbor exchange terms, which seemed unrealistic from a condensed matter perspective - during recent years, the Haldane (and related) models were realized experimentally using photon-dressed graphene~\cite{wang2013}, arrays of coupled
waveguides~\cite{Rechtsman2013} and  periodically modulated optical lattices~\cite{Jotzu2014}. 
Again, it is of particular interest to understand the interaction effects in this model~\cite{Varney2010, Varney2011}. For the case of bosonic particles in the Haldane model, 
He {\it et al.}~\cite{He2015} have recently shown the emergence of a symmetry-protected bosonic integer quantum Hall phase 
by means of numerical simulations of large scale cylinders. In Refs.~\cite{vasic2015,plekhanov2017},  unconventional bosonic chiral superfluid phases have been found.

\begin{figure}[bt]
\centering
\includegraphics[width=1\linewidth]{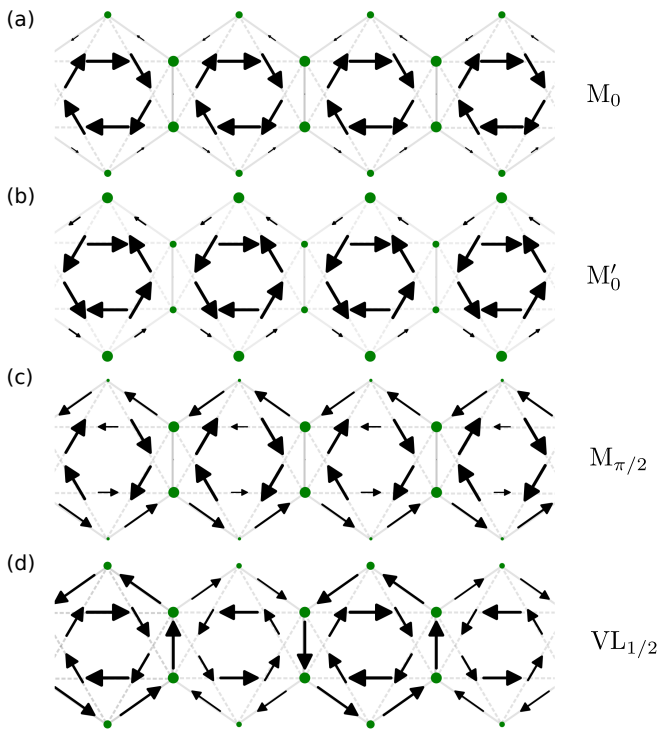}
\caption{(Color online) Examples of the different ground-state current configuration in (a)-(c) the three Meissner-like phases found for fermions and bosons and (d) the vortex-lattice phase that exists for strongly interacting bosons. The data are obtained from actual DMRG simulations for the case of hardcore bosons with  a 
filling $\rho=0.19$, $\phi_H/\pi=0.95$ and (a) $J_H=0.12 J$ (M$_0$), (b) $J_H=0.52 J$ (\Mzp), (c) $J_H=1.4 J$ (\Mpi), (d)  $J_H=1.12 J$ (\VL).
The size of the arrows is proportional to the strength of the currents on the corresponding bond (for clarity, the size of the next-nearest 
neighbor currents has been reduced by a factor of $2$). The size of the circles corresponds to the local onsite density.
{For the depicted bosonic case, the lengths of the largest arrows correspond to currents with an amplitude of (a) $3\cdot 10^{-3} J$, (b) $0.01 J$, (c) $0.04 J$, and (d) $0.16 J$.}
}
\label{fig:patterns}
\end{figure}

\begin{figure*}[tb]
\centering
\includegraphics[width=0.24\linewidth]{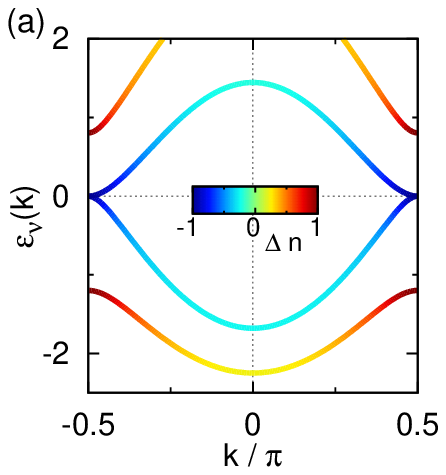}
\includegraphics[width=0.24\linewidth]{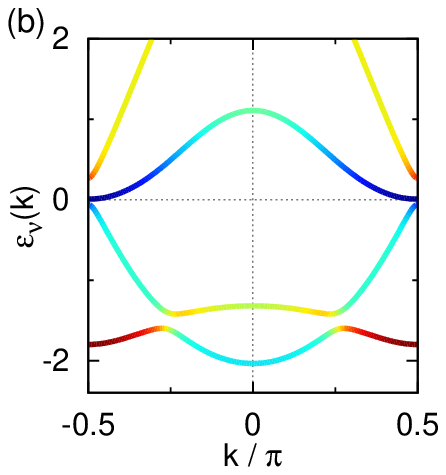}
\includegraphics[width=0.24\linewidth]{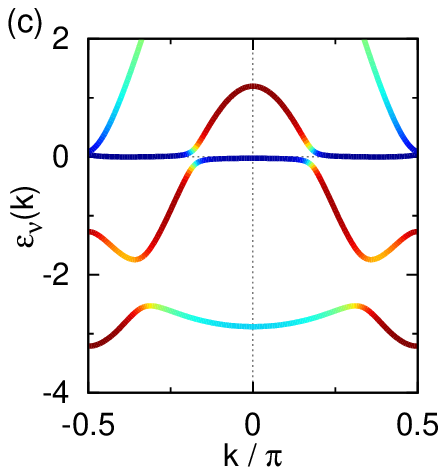}
\includegraphics[width=0.24\linewidth]{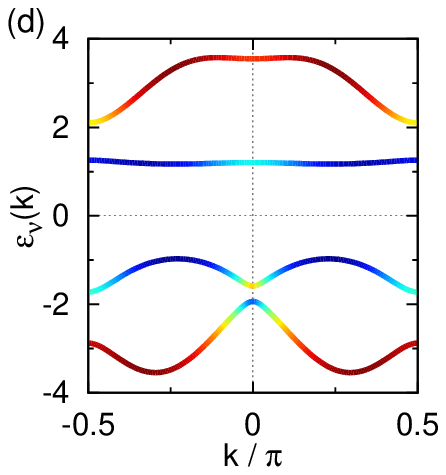}
\caption{(Color online) Examples for different single-particle dispersions $\epsilon_\nu(k)$ realized for  the two-leg Haldane ladder. In the limit 
of low fillings, these correspond to the three Meissner phases, (a) \Mz phase ($\phi_H = 0.95 \pi$, $J_H = 0.1 J$),  (b) \Mzp phase ($\phi_H = 0.95 \pi$, $J_H = 0.4 J$),  (c) \Mpi phase ($\phi_H = 0.95 \pi$, $J_H = 1.1 J$) and (d) the vortex-fluid phase (V) ($\phi_H = 0.6 \pi$, $J_H = J$). The color code depicts the density imbalance $\Delta n$ in the single-particle eigenstates (see Eq.~\eqref{eq:def_dn}). 
Note that in the convention of  Eq.~\eqref{eq:Hk}, the first Brillouin zone is defined as $-\pi/2< k\leq \pi/2$.
}
\label{fig:sf_disp}
\end{figure*}

\begin{table}[tb]
\begin{tabular}{ccccccccc}
\hline
\hline
\hspace{1.cm} & \hspace{0.2cm}$j_A$\hspace{0.2cm} & \hspace{0.2cm}$j_B$\hspace{0.2cm}& \hspace{0.2cm}$\Delta n$\hspace{0.2cm} &  \hspace{0.2cm}$c$\hspace{0.2cm} & $q$ & \hspace{0.2cm}$j_R^{\rm avg}$& \hspace{0.2cm}$\mathcal{O}_{\rm DW}$\hspace{0.2cm}& \hspace{0.2cm}$k_{\rm max}$\hspace{0.2cm}\\[0.1cm]
\hline
\Mz & $\approx 0$ & $\lesssim 0$ & $\gtrsim 0$ & $1$ & $1$ & $0$ & $0$ & $0$\\
\Mzp & $\lesssim 0$ & $\gtrsim 0$ & $\lesssim 0$ & $1$ & $1$ & $0$ & $0$ & $0$\\
\Mpi & $\gtrsim 0$ & $\lesssim 0$ & $\gtrsim 0$ & $1$ & $1$ & $0$ & $0$ & $\pi/2$\\
\hline
V & & & & $2$ & $1$ & $0$ & $0$ & $\pm Q$\\
\hline
\VL & & & & $1$ & $2$ & $>0$ & $>0$ & $0, \pi/2$\\
\hline
\hline
\end{tabular}
\caption{\label{tab:phases} Quantum phases of strongly interacting bosons on the two-leg Haldane ladder studied  in this work. 
The three different Meissner phases \Mz, \Mzp, \Mpi, the vortex-liquid (V) and the vortex lattice \VL phases exist either atop superfluid (SF) or Mott-insulating (MI) states - for simplicity here we just list the properties of the SF phases. We also list  characteristic properties (see the 
text for details) such as the central charge $c$, counting the number of gapless modes, the size $q$ of the effective unit cell in the groundstate (i.e. the number of hexagons), the average local rung current $j_R^{\rm avg}$ (Eq.~\eqref{eq:jRavg}) and the charge-density order $\mathcal{O}_{\rm DW}$ (Eq.~\eqref{eq:ODW}) in the thermodynamic limit. The statements corresponding to the average currents  $j_A$ (Eq.~\eqref{eq:jA}), $j_B$ (Eq.~\eqref{eq:jB}) and the average density difference between $A$ and $B$ sites $\Delta n$ (Eq.~\eqref{eq:def_dn}) for the three different Meissner-phases should be understood as a heuristically observed tendency. $k_{\rm max}$ denotes the position of the largest maximum of the momentum distribution function $n(k)$ (Eq.~\eqref{eq:mom}). For the vortex phase, this typically corresponds to some incommensurate value $0<Q<\pi/2$.
}
\end{table}

An important link between theory and the experimental realization of quantum-lattice gases with artificial gauge fields in the strongly correlated regime can be established by a reduction of the geometry from a two-dimensional  model (which is typically theoretically challenging) to a two-(or multi-)leg ladder system. These quasi-one dimensional models not only allow for an advanced theoretical treatment by means of powerful density matrix renormalization group methods (DMRG)~\cite{White1992, Schollwoeck2011} or analytical bosonization techniques \cite{Giamarchi} but from the experimental perspective, they can be realized using  various different implementations. Besides the superlattice method \cite{Atala2014} and the use of digital mirror devices \cite{Tai2016}, various synthetic-lattice dimension approaches~\cite{Celi2014, Stuhl2015, Mancini2015,Livi2016,Kolkowitz2016,Alex2017} 
 have been employed. These use a coupling between internal states to
realize some or even all lattice directions.
While the theoretical interest in ladders with a flux dates back to early studies of Josephson junction arrays \cite{Kardar1986,Granato1990,Mazo1995,Denniston1995}, 
which was then extended to the strongly interacting regime in a seminal paper by Orignac and Giamarchi \cite{Orignac2001},
the prospects of experimental realizations with ultracold quantum gases have led to tremendous theoretical activity.  
In particular, during the past years, the study of the low-dimensional relatives of, for example, the Hofstadter-Harper model on two- or three-leg ladder geometries have attracted a large deal of interest~\cite{Orignac2001,Carr2006,Roux2007,Petrescu2013,Tokuno2014,Huegel2014,Wei2014, Uchino2015,Barbarino2015,Zeng2015,Petrescu2015,Cornfeld2015,Ghosh2015,Yan2015,Piraud2015,DiDio2015,Kolley2015,Natu2015,Petrescu2015,Keles2015,Taddia2016,Uchino2016,Ghosh2016,Anisimovas2016,Greschner2016,Orignac2017,Petrescu2017,Strinati2017}.
While fermionic systems are equally interesting \cite{Carr2006,Roux2007}, much work has focussed on the ground-state phase diagram of {\it bosonic} systems, observing a multitude of phases resulting from the kinetic frustration due to the presence of a homogeneous flux per plaquette.
These include three Meissner phases characterized by a uniform edge current as well as commensurate and incommensurate vortex-fluid phases \cite{Orignac2001,Petrescu2013,Piraud2015}. 
These phases can be characterized by the behavior of the chiral edge current and bulk currents or are distinct by the spontaneous breaking of  a discrete symmetry (see Ref.~\cite{Greschner2016} for an overview).
We will refer to the ladders that result from the thin-cylinder limit of the Hofstadter-Harper model as flux ladders. Recent work addresses the possibility of stabilizing low-dimensional relatives of fractional quantum Hall states
in ladder systems \cite{Cornfeld2015,Grusdt2014,Petrescu2015,Petrescu2017,Strinati2017, Greschner2017}.

In this paper we study the ground-state physics of the bosonic Haldane model on a two-leg ladder geometry, which exhibits a  rich physics.
 We will focus our analysis on  the low filling regime of less than one particle per unit-cell.
We start our analysis with a description of the free-fermion version of the model~\eqref{eq:HH}, which allows us to understand some of the ground-state phases, and compare to the properties of hardcore bosons ($(c_{\ell}^\dagger)^2=0$). These include Meissner-like states and an incommensurate vortex-fluid phase. The most striking difference compared to free fermions is the emergence of a phase with a broken translational symmetry and the effectively doubled unit-cell in the hardcore boson limit, which we will call vortex-lattice phase \VL in analogy to the phases studied for flux ladders~\cite{Orignac2001, Greschner2016}. Figure~\ref{fig:patterns} shows representative density and current configurations for the main quantum phases described in this work. 
By means of DMRG simulations and weak-coupling methods we study the formation of the \VL phase for finite onsite interactions
$U<\infty$ as well, where the Hamiltonian is augmented by the term
\begin{equation}
H_{\rm int}= \frac{U}{2} \sum_{\ell} n_{\ell} (n_{\ell}-1)\,, 
\end{equation}
with $n_{\ell}=c_\ell^\dagger c_\ell $.
Some exact-diagonalization results for a similar ladder variant of this model have been discussed in \cite{plekhanov2017}.

The paper is organized as follows. We start our discussion of the Haldane ladder from the single-particle perspective presented in Sec.~\ref{sec:free}. 
For the case of free fermions, we introduce the basic properties of the different Meissner-like phases of the model and define relevant observables. In order to give a specific example, we will first fix the phase $\phi_H$ to be close to $\pi$. {Due to symmetry, the case $\phi_H=\pi$ does not exhibit  finite local currents and we therefore   choose $\phi_H=0.95\pi$ unless stated otherwise.} We  study the properties and analyze the ground-state phase diagram as a function of the next-nearest-neighbor 
tunneling amplitude $J_H$. In the following sections Secs.~\ref{sec:vl}-\ref{sec:phiH},  we study the behavior of the bosonic model for the same range of parameters starting from the case of hardcore bosons. Here, we focus on the properties of the \VL phase, which is one of the main results of this paper. For the case of finite interactions and in Sec.~\ref{sec:U}, we develop a weak-coupling picture of the emergence of the \VL phase, which we compare to numerical simulations. Finally, we  discuss the ground-state phase diagram of hardcore bosons as a function of the phase $\phi_H$ for a fixed amplitude $J_H$ in Sec.~\ref{sec:phiH} and conclude with a brief summary of our results presented in Sec.~\ref{sec:sum}.

\section{Single-particle spectrum and free fermions}
\label{sec:free}

We start our analysis of the model from the free-fermion limit, which allows us to derive an initial picture of some of the liquid phases found also for bosons.

We express the Hamiltonian~\eqref{eq:HH} in momentum space as $H_H = \sum_k \vec{\tilde{c}}_k^\dagger \mathcal{H}_H(k) \vec{\tilde{c}}_k$. Here the momentum-space representations of the annihilation operators of the unit cell are grouped into a single vector $\vec{\tilde{c}}_k$ with $\vec{\tilde{c}}_k = \sum_{\ell} \e^{2 \ii k} \vec{c}_{\ell}$  and $\vec{c}_{\ell}^T = \left( c_{A_{1,\ell}}, c_{A_{2,\ell}}, c_{B_{1,\ell}}, c_{B_{1,\ell}} \right)$ and
\begin{widetext}
\begin{align}
\mathcal{H}_H(k) = \left(
\begin{array}{cccc}
 2 J_H \cos (2k+\phi_H) & J & \left(1+\e^{-2 \ii k}\right) J &  2 J_H \e^{-\ii k} \cos \left(k-\phi_H\right) \\
 J & 2 J_H \cos (2k-\phi_H) & 2 J_H  \e^{\ii k} \cos \left(k+\phi_H\right) & \left(1+\e^{-2 \ii k}\right) J \\
 \left(1+\e^{2\ii k}\right) J & 2 J_H  \e^{-\ii k} \cos \left(k+\phi_H\right) & 0 & 0 \\
2 J_H \e^{\ii k} \cos \left(k-\phi_H\right) & \left(1+\e^{2 \ii k}\right) J & 0 & 0 \\
\end{array}
\right)\,.
\label{eq:Hk}
\end{align}
\end{widetext}
This can readily be diagonalized leading to $H_H = \sum_{k,\nu=1,\dots, 4} \epsilon_\nu(k) \alpha^{\dagger}_{\nu k} \alpha_{\nu k}$ with new operators $\alpha_{\nu k} = \sum_\gamma \mathcal{U}_{\gamma, k} c_{\gamma,k}$ living on four generally separated energy bands $\epsilon_\nu(k)$ with band index $\nu=1,2,3,4$.

For $\phi_H\approx\pi$, in particular, we find a rich bandstructure $\epsilon_\nu(k)$. For concreteness and unless stated otherwise, we fix the value of the  phase to $\phi_H=0.95\pi$. 
Since this is slightly detuned from $\phi_H=\pi$, there are  finite chiral currents.  We then vary the ratio $J_H/J$ as a free parameter.

Figure~\ref{fig:sf_disp} shows four examples of the single-particle spectrum for various values of $J_H/J$ and $\phi_H$ for which different kinds of lowest-band minima are realized: a single minimum at $k=0$ (Figs.~\ref{fig:sf_disp}~(a) and (b)), a single minimum at $k=\pm\pi/2$ (Fig.~\ref{fig:sf_disp}~(c)) or two degenerate minima at $k=\pm Q$ (Fig.~\ref{fig:sf_disp}~(d)). We may associate these situations to four different low-density ground-state phases - three Meissner-like phases \Mz, \Mzp, \Mpi and an incommensurate vortex-fluid phase (V),  which we will discuss in the following.

We may best characterize the different phases  by calculating  their local current and density configurations. Due to the explicitly broken time-reversal symmetry of the Hamiltonian~\eqref{eq:HH}, quantities of interest are the typically non-vanishing local and average particle currents  on nearest-and next-nearest neighbor bonds. 
A local current $\mathcal{J}(\ell\to \ell')$ from site $\ell$ to site $\ell'$ can be derived from  the continuity equation $\langle \partial n_{\ell} / \partial t \rangle = - \sum_{\langle \ell \ell' \rangle} \mathcal{J}(\ell \to \ell') - \sum_{\langle\langle \ell \ell' \rangle\rangle} \mathcal{J}(\ell \to \ell')$.

Examples of the different local current structures in the three Meissner-like states are shown in Figs.~\ref{fig:patterns}(a), (b) and (c). Although the data shown are computed for  hardcore bosons, the corresponding low-filling free-fermion version of these states looks similar. 
Since the nearest-neighbor currents $\mathcal{J}(A_{\gamma,\ell}\to B_{\gamma,\ell})$ and $\mathcal{J}(B_{\gamma,\ell} \to A_{\gamma,\ell+1})$ 
flow in the same direction  along the legs, we  dub the three phases Meissner phase (M). The inner currents on the rungs $\mathcal{J}(A_{1,\ell}\to A_{2,\ell})$ 
are strongly suppressed.

\begin{figure}[tb]
\centering
\includegraphics[width=1\linewidth]{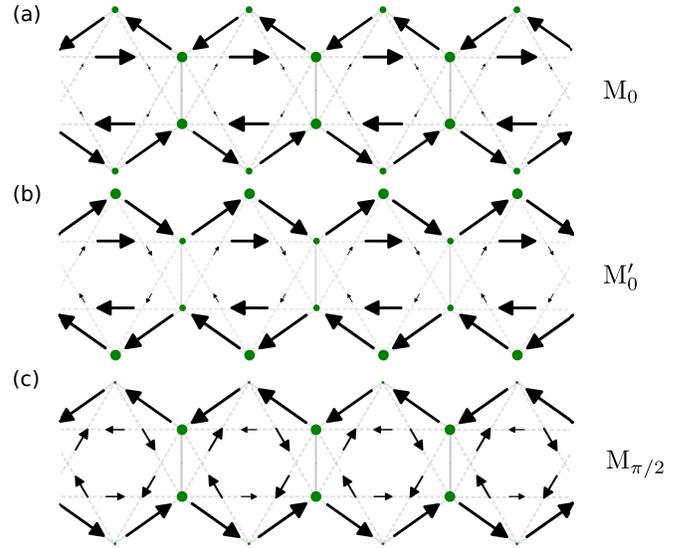}
\caption{(Color online) The same current configurations as in Fig.~\ref{fig:patterns}, but the position of $A$ sites has been exchanged, i.e., $A_{1,\ell} \leftrightarrow A_{2,\ell}$. In this way, the Meissner character of the phases (a)-(c) becomes more evident since the strongest currents now flow through the outer boundary of the ladder system.}
\label{fig:inv_patterns}
\end{figure}

In order to make the ``Meissner character'' of the phases more evident, in Fig.~\ref{fig:inv_patterns}, we display the same current configurations for the three different Meissner phases of Fig.~\ref{fig:patterns} with swapped positions of the $A$ sites, i.e., relabeling $A_{1,\ell} \leftrightarrow A_{2,\ell}$. In this notation the strongest current runs through the outer boundary of the ladder system, which is a characteristic signature of  a Meissner phase \cite{Piraud2015}.

We may define an average chiral current on the nearest-neighbor bonds as 
\begin{align}
\jc = -\frac{1}{L} \sum_{\ell} \lbrack \mathcal{J}(A_{\gamma,\ell}\to B_{\gamma,\ell}) + \mathcal{J}(B_{\gamma,\ell} \to A_{\gamma,\ell+1})  \rbrack\,.
\end{align}
In order to take into account the inner currents running on the next-nearest neighbor bonds, we also introduce the average current $j_A$ ($j_B$) that runs through an $A_1$ ($B_1$) site: 
\begin{align}
j_A &= -\frac{1}{L} \sum_{\ell}  \left \lbrack \mathcal{J}(A_{1,\ell}\to B_{1,\ell}) + \mathcal{J}(A_{1,\ell}\to A_{1,\ell+1})  \right. \nonumber \\
& \left. + \mathcal{J}(A_{1,\ell}\to B_{2,\ell}) + \mathcal{J}(A_{1,\ell}\to A_{2,\ell}) \right\rbrack
\label{eq:jA}
\end{align}
and
\begin{align}
j_B &= -\frac{1}{L} \sum_{\ell} \left \lbrack \mathcal{J}(B_{1,\ell}\to A_{1,\ell+1}) + \mathcal{J}(B_{1,\ell}\to A_{2,\ell+1}) \right\rbrack\,.
\label{eq:jB}
\end{align}
We can understand $j_A$ 
as an observable that quantifies the average current that runs from one hexagon to the neighboring one, while $j_B$ quantifies the current circulating inside the hexagon.

In  the four-site unit cell, the density difference between $A$ and $B$ sites,
\begin{align}
\Delta n = \frac{1}{L} \sum_{\ell,\gamma=1,2} (n_{A_{\gamma,\ell}} - n_{B_{\gamma,\ell}})\,  \,,
\label{eq:def_dn}
\end{align}
typically is nonzero. We will refer to $\Delta n$ as the density imbalance. 

\begin{figure}[tb]
\centering
\includegraphics[width=1.0\linewidth]{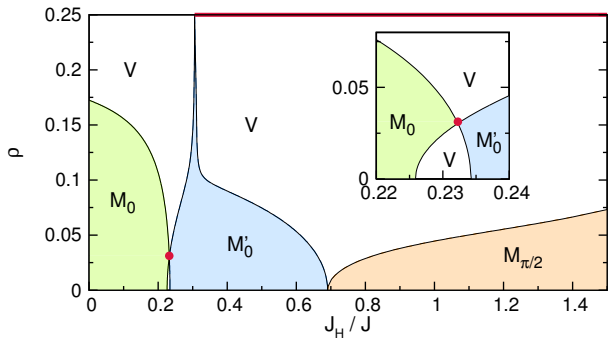}
\caption{(Color online) Phase diagram for free spinless fermions in the Haldane ladder ($\phi_H=0.95 \pi$) as a function of density and $J_H/J$. Three Meissner-like phases exist (\Mz, \Mzp, \Mpi) as well as a vortex fluid phase (V) with incommensurate vortex density. The large red circles  mark a Dirac-like point, for which the dispersion relation remains linear.}
\label{fig:sf_pd}
\end{figure}

\begin{figure}[tb]
\includegraphics[width=1.\linewidth]{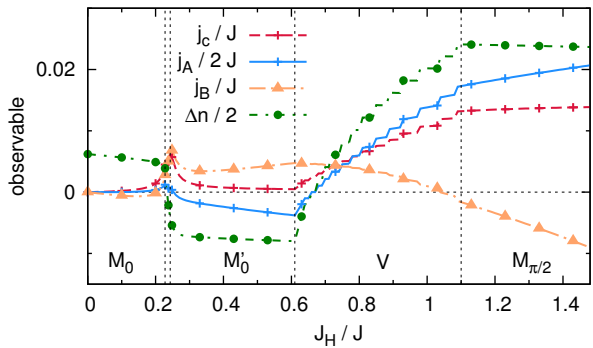}
\caption{(Color online) Cut through the phase diagram of free fermions Fig.~\ref{fig:sf_pd} at a small filling $\rho=0.05$, showing
 the currents $\jc$, $j_A/2$, $j_B$, and the density imbalance $\Delta n/2$ for a system of $L=159$ sites. Note that for clarity, less symbols than available data points are shown.
Vertical lines indicate the phase boundaries extracted from the bandstructure.}
\label{fig:sf_TH_cut}
\end{figure}

In Fig.~\ref{fig:sf_pd}, we show the ground-state phase diagram of free fermions as a function of filling $\rho=N/L$ (up to one particle per unit-cell) and the nearest-neighbor tunneling amplitude $J_H/J$ with extended regions of the \Mz, \Mzp and \Mpi phases. 
In Fig.~\ref{fig:sf_TH_cut}, the current and the density imbalance for a cut through the phase diagram at low filling are depicted.
In the \Mz phase, the local currents on the next-nearest-neighbor bonds all circulate in the  clockwise direction, opposite to the (small) currents on the nearest-neighbor links. Due to this almost closed ring-current within the hexagon, $j_A$ approximately vanishes in this phase. 
In the \Mzp phase, the sign of the diagonal next-nearest-neighbor currents is flipped compared to the \Mz phase. Hence, also the sign of $j_B$ is inverted compared to the \Mz phase and we observe a finite inter-hexagon current $j_A<0$.
While in both the \Mz and \Mzp phases the chiral current on the outer nearest neighbor bonds $j_c$ is strongly suppressed, the \Mpi phase is characterized by a larger $j_c$. Furthermore we find $j_B<0$ and $j_A>0$ opposite to the \Mzp phase.

The expectation value of $\Delta n$ may be used to further distinguish the \Mz and \Mzp phases from each other (as can be also inferred from the color-code of the dispersion-relations in Fig.~\ref{fig:sf_disp}). The \Mz and \Mpi phases have $\Delta n >0$, while $\Delta n <0$ for the \Mzp phase.

For higher fillings (or for special parameters also in the dilute limit) one encounters the situation that more than one Fermi-sea forms, either by occupying modes of an overlapping higher band or because a second local minimum of the same band gets occupied. The doubling of the number of Fermi-points is reflected by a change of the central charge parameter from $c=1$ to $c=2$.
Due to the correspondence with the flux-ladder case~\cite{Orignac2001, Piraud2015}, we generally refer to these phases  as vortex fluid phases (V) since the local current structure for a system with open boundary conditions exhibits a strong oscillatory but incommensurate pattern (see Sec.~\ref{sec:phiH} for a discussion of the analogous vortex-fluid phase for the case of hardcore bosons).
Interestingly, for the parameters of Fig.~\ref{fig:sf_pd} and at the crossing from the  \Mz to the \Mzp  phase, a tiny region with a doubly-degenerate lowest band minimum emerges
(see the inset in Fig.~\ref{fig:sf_pd}).

For special parameters, Dirac-like points exist, in which two bands touch with a linear dispersion relation  (see the red dot in Fig.~\ref{fig:sf_pd}). While for interacting fermions, nontrivial effects might be expected, for the case of (interacting) bosons this feature plays no role since finite-filling properties do not carry over from fermions to bosons. For the filling of one particle per unit cell a trivial band-insulating state is realized for larger values of $J_H/J$.

\section{Ground-state phase diagram for hardcore bosons}
\label{sec:vl}

In the following we move on to the case of an interacting, single-component gas of bosons on the Haldane ladder.  We start with the  case of hardcore bosons ($U/J\to\infty$),  which is the simplest case from the  numerical perspective due to its restricted local Hilbert-space and provides a good starting point to investigate the effect of interactions.

\subsection{Diagnostic tools}
Since this model is no longer exactly solvable,  we perform density matrix renormalization group  (DMRG) simulations \cite{White1992,Schollwoeck2005,Schollwoeck2011} with open boundary conditions to study the ground-state physics of this model keeping up to $\chi=1000$ DMRG states. We consider various system sizes of odd numbers of rungs $L$, such that we simulate systems with $(L-1)/2$ hexagons.

Apart from extracting various order parameters, the density imbalance and local currents, our DMRG calculations allow us to study further interesting quantum-information measures. For example, the block entanglement entropy $S_{vN} = -{\rm Tr}\lbrack  \rho_{l} \log \rho_l \rbrack $, 
for the reduced density matrix $\rho_l$ of a subsystem of length $l$ may be employed to extract the central charge from the so-called Calabrese-Cardy formula~\cite{Holzhey1994, Vidal2003, Korepin2004, Calabrese2004}
\begin{align}
S_{vN} = \frac{c}{3} \log\left[ \frac{L}{\pi} \sin \frac{\pi l}{L} \right] + \cdots\,.
\label{eq:CCformula}
\end{align}
Phase transitions may also be detected in the finite-size scaling of the fidelity susceptibility~\cite{gu2010}
\begin{align}
\chi_{FS}(J_H) = \lim_{\delta J_H\to 0} \frac{-2 \ln |\langle \Psi_0(J_H) | \Psi_0(J_H + \delta J_H) \rangle| }{(\delta J_H)^2} \,,
\end{align}
with $|\Psi_0\rangle$ being the ground-state wave-function.

\subsection{Phase diagram}

\begin{figure}[tb]
\centering
\includegraphics[width=1.0\linewidth]{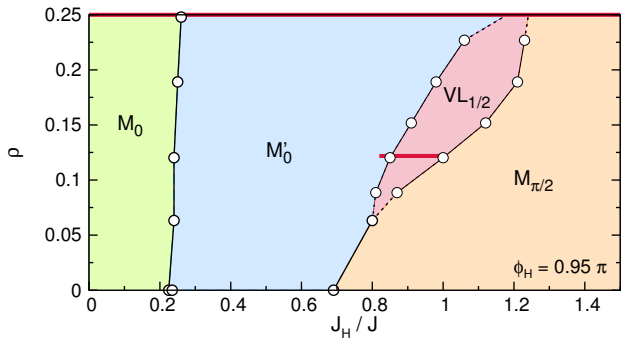}
\caption{(Color online) Phase diagram for hardcore bosons in the Haldane ladder for $\phi_H=0.95\pi$ as a function of density $\rho$ and the next-nearest-neighbor 
hopping parameter $J_H/J$. There are three Meissner-like phases (\Mz, \Mzp, \Mpi) as well as a vortex-lattice phase \VL. At density $\rho=1/8$, a Mott-insulating state 
exists in the range $ 0.75 \lesssim J_H/J \lesssim 1$, while at $\rho=1/4$, the system is in a Mott-insulating state for any $J_H/J$. }
\label{fig:pdH2}
\end{figure}

In Fig.~\ref{fig:pdH2} the ground-state phase diagram of hardcore bosons is shown for the  parameters of Fig.~\ref{fig:sf_pd}.
 While in the  limit of a dilute lattice gas, the same sequence of ground-state phases as for the case of free fermions is observed, for larger fillings,  the differences become more drastic since the incommensurate vortex-fluid phases are suppressed while a vortex-lattice (\VL) phase gets stabilized, which we will describe in the following Sec.~\ref{sec:VL} in more detail.
The current and density structure of this unconventional \VL phase is shown in Fig.~\ref{fig:patterns}(d). 

\begin{figure}[tb]
\includegraphics[width=1.\linewidth]{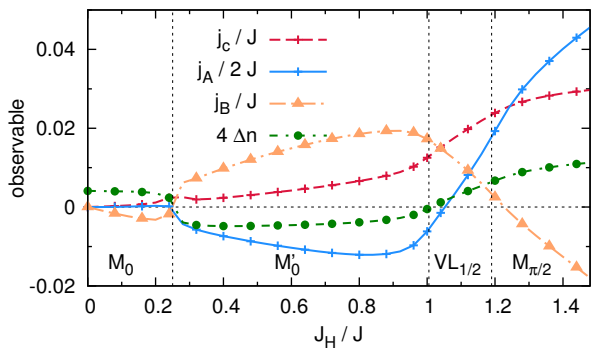}
\caption{(Color online) Cut through the phase diagram of hardcore bosons Fig.~\ref{fig:pdH2} at density $\rho=0.19$, showing the currents
$j_c$, $j_A/2$, $j_B$, and the density imbalance $4\Delta n$.
}
\label{fig:TH_cut_jc}
\end{figure}

In Fig.~\ref{fig:TH_cut_jc}, we show several observables and chiral currents for a cut through the phase diagram at a fixed density. As already anticipated in the previous section, the three different Meissner-like phases \Mz, \Mzp and \Mpi show a  behavior similar to the free-fermion case discussed earlier (see Fig.~\ref{fig:sf_TH_cut}): \Mz and \Mzp phases can be discriminated by the sign change of the $\Delta n$ and $j_B$ observables. While the \Mz phase is characterized by $j_A\approx 0$ and $j_B<0$, we observe $j_A<0$ and $j_B>0$ in the \Mzp phase and opposite signs, $j_A>0$ and $j_B<0$, in the \Mpi phase.
By means of our numerical simulations we cannot resolve any intermediate phase between the \Mz and \Mzp phases at finite densities.

For certain commensurate fillings, namely at $\rho=1/4$ and for the various Meissner phases but also at $\rho=1/8$ for the \VL phase, a charge gap opens (horizontal thick line in Fig.~\ref{fig:pdH2}). 
With this, the sequence of phase transitions becomes very rich, since the \Mz, \Mzp and \Mpi phases may be observed on both a superfluid (SF) and a gapped Mott-insulator (MI) background.
Contrary to the free-fermion case, we observe the opening of a charge gap at $\rho=1/4$ for all values of $J_H/J>0$. At filling $\rho=1/8$ the MI-phase is apparently confined to the region of the \VL phase. From our calculations, we cannot exclude the possibility of a small surrounding region of \Mz-MI and \Mpi-MI phases at filling $\rho=1/8$. Further details of the gapped regions will be discussed below in Sec.~\ref{sec:phiH}.

\begin{figure}[tb]
\includegraphics[width=1.\linewidth]{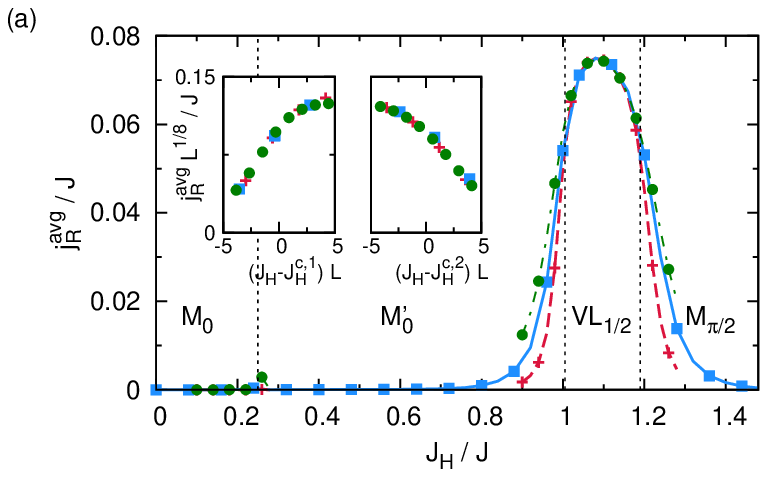}
\includegraphics[width=1.\linewidth]{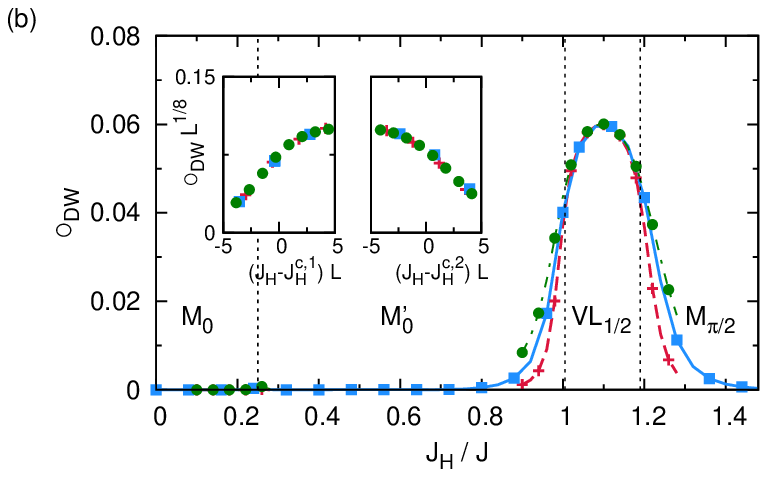}
\caption{(Color online) Order parameters for the \VL phase for a cut through the phase diagram of hardcore bosons Fig.~\ref{fig:pdH2} at density $\rho=0.19$:
  (a) Average rung-current $j_R^{\rm avg}$ and (b) charge-density order parameter $\mathcal{O}_{\rm DW}$.
The insets depict the collapse of the  finite system-size data onto 
one curve close to the phase-transition points assuming an Ising-type scaling relation. 
 We find $J_H^{c,1}\approx J$ and $J_H^{c,2}\approx 1.19 J$ for the left and right boundary of the \VL phase, respectively.
}
\label{fig:TH_cut}
\end{figure}

\begin{figure}[tb]
\includegraphics[width=1.\linewidth]{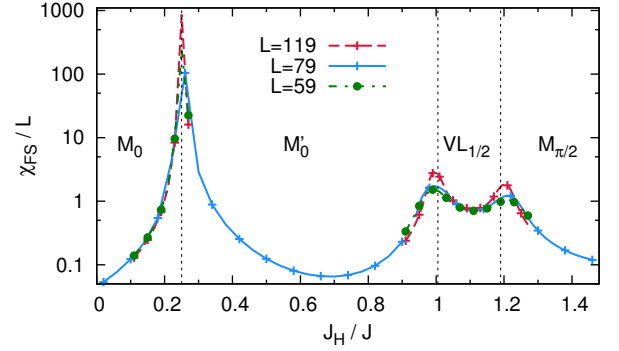}
\caption{(Color online) Scaling of the fidelity susceptibility $\chi_{FS}(J_H)/L$ for a cut through the phase diagram of hardcore bosons Fig.~\ref{fig:pdH2} at density $\rho=0.19$. Three phase transitions can be observed at which $\chi_{FS}(J_H)/L$ diverges.}
\label{fig:TH_cut_fs}
\end{figure}

\begin{figure}[tb]
\includegraphics[width=1.0\linewidth]{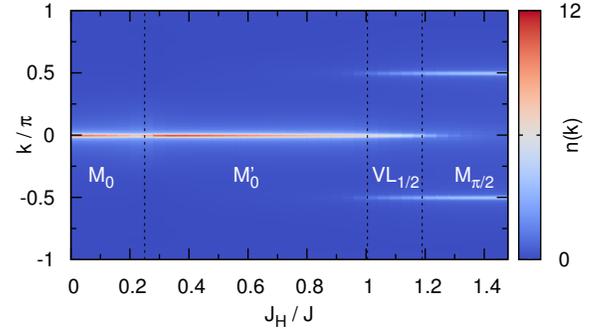}
\caption{(Color online) Quasimomentum distribution function $n(k)$ on the legs for the cut through the phase diagram of hardcore bosons Fig.~\ref{fig:pdH2} at $\rho=0.19$. The \VL phase is characterized by three distinct peaks at $k=0$ and $\pm \pi/2$.}
\label{fig:TH_mom}
\end{figure}

\subsection{Vortex-lattice phase}
\label{sec:VL}

Contrary to the vortex-fluid phases, the \VL-SF phase  is a single-component phase with a central charge $c=1$ and it exhibits a spontaneously broken translational and parity symmetry. Therefore, the effective unit-cell is doubled as can be seen in Fig.~\ref{fig:patterns}(d). 
An order parameter  
for the \VL phase can be defined from its average local rung-current
\begin{align}
j_R^{\rm avg} = \frac{1}{L} \sum_{\ell=0,\dots, L} \left| \mathcal{J}(A_{\ell,1}\to A_{\ell,2}) \right| \,.
\label{eq:jRavg}
\end{align}
As an example, we present the   finite average rung-current $j_R^{\rm avg}$ in this region in Fig.~\ref{fig:TH_cut}(a). As shown in the insets of Fig.~\ref{fig:TH_cut}(a) the scaling of 
 $j_R^{\rm avg}$
close to the quantum critical points follows Ising-scaling relations and the data points from several finite system-size simulations can be collapsed onto one single curve.
As can be seen from the local density pattern shown in Fig.~\ref{fig:patterns}(d), the \VL phase exhibits a finite density oscillation between adjacent unit-cells. Hence, we may define a charge-density-wave order parameter via
\begin{align}
\mathcal{O}_{\rm DW} = \frac{1}{L} \left| \sum_{\ell=0,\dots, L,\gamma} (-1)^{\ell} (n_{A_{\ell,\gamma}} + n_{B_{\ell,\gamma}}) \right| \,.
\label{eq:ODW}
\end{align}
Our numerical calculations indicate that $\mathcal{O}_{\rm DW}$ stays finite in the thermodynamic limit (see the data for  $\mathcal{O}_{\rm DW}(L)$ shown in Fig.~\ref{fig:TH_cut}(b)), and its $L$-dependence indeed looks almost identical to the plot of $j_R^{\rm avg}(L)$ in Fig.~\ref{fig:TH_cut}(a).

A further indication  of the Ising character of both the \Mzp to \VL as well as the \VL to \Mpi transition is the approximate linear divergence  of the peak of the fidelity susceptibility $\chi_{FS}/L \propto L$ with  system size $L$ as seen in Fig.~\ref{fig:TH_cut_fs}~\cite{Venuti2007,Greschner2013b}.
Contrary to that, the highly non-linear scaling of the maximum of $\chi_{FS}(J_H)/L$ close to the \Mz to \Mzp transition with respect to  system size $L$  (see Fig.~\ref{fig:TH_cut}~(a)) may indicate a first-order transition. The same appears to be the case for the \Mzp-to-\Mpi transition at low fillings. However, due to the finite resolution of our calculations, we cannot exclude the possibility of small intermediate phases.

The quasimomentum distribution function
\begin{align}
n(k)=\frac{1}{L}\sum_{\ell,\ell'}{e^{\ii k(\ell-\ell')}\langle{a^{\dagger}_{\ell}a_{\ell'}}\rangle}
\label{eq:mom}
\end{align}
is particularly interesting as a possible experimental signature of the \VL phase. 
As shown in Fig.~\ref{fig:TH_mom}, the \VL-SF phase is characterized by sharp peaks in the quasimomentum distribution at $k=0$ and $k=\pm \pi/2$.

{At incommensurate fillings the \VL phase does not exhibit a charge gap and the single-particle correlations decay algebraically (as can  also be seen from the presence of sharp peaks in Fig.~\ref{fig:TH_mom}). We may hence understand this liquid \VL-SF phase with charge-density ordering as another type of a lattice supersolid phase~\cite{andreev1969,Prokof2005,Batrouni2006,pollet2010,Landig2016}.}

The \VL phase may be seen as an analog of the vortex-lattice phase known from the soft-core boson vortex-lattice phases on flux ladders~\cite{Orignac2001, Greschner2016}, or the so-called chiral phases known from frustrated zig-zag ladders~\cite{Greschner2013}. We want to stress, however, some important differences:
The flux-ladder vortex-lattice phases are known to be the most stable for the case of weak interactions and are completely suppressed for the case of hardcore particles on two-leg flux ladders~\cite{Piraud2015}. Here, however, we find the vortex lattice phase even for $U/J\to \infty$.

The zig-zag ladder chiral phases, on the other hand,  are best understood from the dilute limit $\rho\to 0$~\cite{Kolezhuk2012} in which it can be connected to the presence of a two-fold degenerate band-minimum for an extended parameter range, where interactions may favour either a two-component phase or a single-component chiral phase with spontaneously broken symmetry between the two dispersion minima. 
A similar mechanism applies to the so-called biased ladder phase (BLP) on two-leg flux ladder systems~\cite{Wei2014,Greschner2016}, which is, however, again most stable for the case of small interactions $U\to 0$.

In the present case of  the \VL phase on the Haldane ladder, the single-particle spectrum is degenerate only for a single point at $J_H=J_H^c$. As we will motivate in the following section, one may understand this \VL phase, naively transferring from the free-fermion case, as a spontaneous breaking of an effective emergent degeneracy between $k=\pi/2$ and $k=0$ modes due to the finite filling and interactions.

\section{Finite interaction strengths $U/J<\infty$}
\label{sec:U}

\begin{figure}[tb]
\centering
\includegraphics[width=1\linewidth]{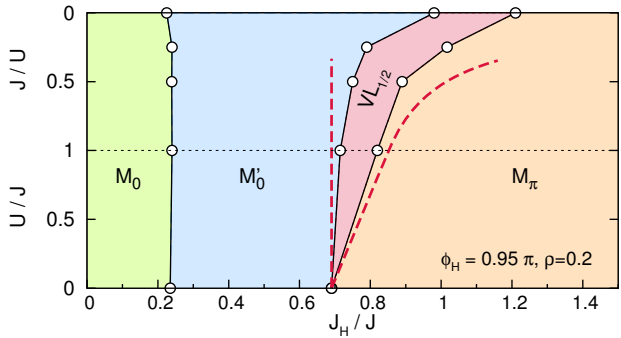}
\caption{(Color online) Phase diagram for soft-core bosons in the Haldane ladder as a function of $J_H/J$ and of the interaction strength $U/J$. Note that for $U>J$, 
we plot the inverse interaction strength $J/U$ in order to connect the phase boundaries to the hardcore-boson limit. The dashed lines denote the region of instability as predicted from the weak-coupling Bogoliubov approximation (see the text in Sec.~\ref{sec:Uweak}).}
\label{fig:pd_fU}
\end{figure}

\begin{figure}[tb]
\centering
\includegraphics[width=0.49\linewidth]{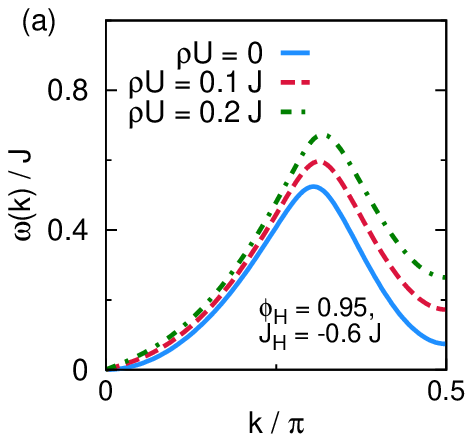}
\includegraphics[width=0.49\linewidth]{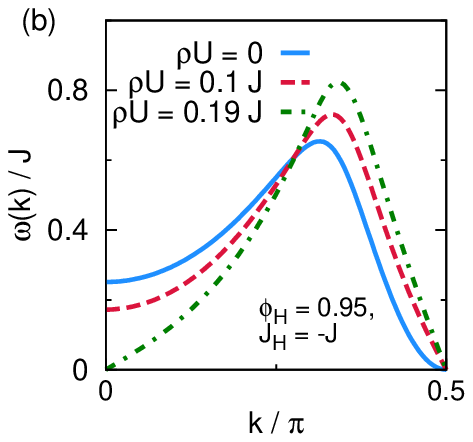}
\caption{(Color online) Bogoliubov excitation spectrum $\omega(k)$ for several interaction strengths $\rho U$ for $\Phi_H=0.95\pi$ and (a) $J_H=0.6J$ and (b) $J_H=J$.
}
\label{fig:bog_e}
\end{figure}

In the following we analyze the stability of the vortex-lattice phase \VL for the case of finite repulsive on-site interactions $U/J<\infty$. The main analytical and numerical results are summarized in the phase diagram of Fig.~\ref{fig:pd_fU}.

\subsection{Limit of weak interactions}
\label{sec:Uweak}

In the weak-interaction limit $\rho U \to 0$, we may shed light on the mechanism for the stabilization of the \VL phase by means of  a simple Bogoliubov-like approximation. We start by projecting the interaction to the lowest band
\begin{align}
H_{\rm eff} = &\sum_k \epsilon_\alpha(k) \alpha_k^\dagger \alpha_k  \nonumber\\
 &+ \frac{U}{2} \sum_{k,k',q} V_{k+q,k'-q, k,k'}\, \alpha_{k+q}^\dagger \alpha_{k'-q}^\dagger \alpha_{k} \alpha_{k}
\end{align}
with $V_{k_1,k_2, k_3, k_4} = \sum_{\nu=1\dots 4}  \mathcal{U}^\dagger_{k_1,\nu} \mathcal{U}^\dagger_{k_2,\nu} \mathcal{U}_{k_3,\nu} \mathcal{U}_{k_4,\nu}$, where
$\mathcal{U}_{k,\nu}$ is the $\nu$-th eigenvector of the Hamiltonian matrix Eq.~\eqref{eq:Hk}.
As an approximation in the limit of $\rho U\to 0$, we assume a quasicondensation of the bosons at $Q=0$ (for $J_H \lesssim J_H^c \approx 0.69\dots J$) or at $Q=\pi/2$ and with $\alpha_Q \approx \sqrt{N} + \tilde{\alpha}_Q$. Using $\alpha_Q^\dagger \alpha_Q^\dagger \alpha_Q \alpha_Q \approx N^2 - 2 N\sum_{k\neq Q} \tilde{\alpha}_k^\dagger \tilde{\alpha}_k$ we rewrite the Hamiltonian retaining only quadratic terms
\begin{align}
H_{\rm eff} \approx \sum_{k} \mathcal{A}(k) \left(\tilde{\alpha}_k^\dagger \tilde{\alpha}_k +\tilde{\alpha}_{-k}^\dagger \tilde{\alpha}_{-k} \right) \nonumber\\
+ \sum_{k} \mathcal{B}(k) \left(\tilde{\alpha}_k^\dagger \tilde{\alpha}^\dagger_{-k} +\tilde{\alpha}_{k} \tilde{\alpha}_{-k} \right)
\end{align}
with 
\begin{align}
\mathcal{A}(k)& =\epsilon(k) - \epsilon(Q) + 4U\rho \left( 2 V_{Q,k,Q,k} - V_{Q,Q,Q,Q}\right) \,,\nonumber\\
\mathcal{B}(k)& = 2U\rho \left( V_{k,-k,Q,Q} +V_{Q,Q,k,-k}\right) \,.
\end{align}
A standard Bogoliubov transformation $\beta_k = u_k \tilde{\alpha}_k - v_k \tilde{\alpha}_{-k}^\dagger$ diagonalizes the effective model $H_{\rm eff} = E_0 + \sum_k \omega(k) \beta_k^\dagger \beta_k$ with $\omega(k) = \sqrt{A(k)^2-B(k)^2}$, where $E_0$ is the ground-state energy.  Examples of the Bogoliubov-excitation spectra $\omega(k)$ for values of $J_H/J$ in the \Mz and the \Mpi phases are shown in Fig.~\ref{fig:bog_e}. 

Starting at $U=0$ from the \Mpi phase, with increasing interaction $\rho U$, the second minimum of the dispersion relation decreases and at some critical value touches zero at $k=0$ as is shown in Fig.~\ref{fig:bog_e}(b). 
At this point the solution becomes instable and the approximation of a single quasicondensate at $Q=\pi/2$ is no longer valid. A finite occupation of modes around $k=0$ has to be taken into account. Hence, we may associate this point of instability with the formation of a phase with a strong interplay between 0 and $\pi/2$ modes, which for large values of $\rho U$ can be identified to be the \VL phase. Note that the 
\VL  phase is characterized by three maxima in the quasimomentum distribution function at $k=0,\pm \pi/2$.
Interestingly, starting from the \Mzp phase, 
the quasicondensate at $Q=0$ seems to be stabilized with increasing $U\rho$ and the second local minimum at $k=0$ vanishes upon increasing the interaction strength, as shown in  Fig.~\ref{fig:bog_e}(a).

\begin{figure}[tb]
\centering
\includegraphics[width=1.\linewidth]{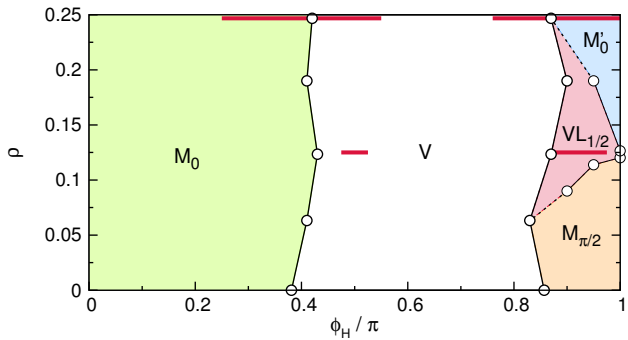}
\caption{(Color online) Phase diagram for hardcore-bosons in the Haldane ladder as a function of the phase $\phi_H$ and the filling $\rho$ for $J_H=J$.
At $\rho=1/4$, there are two regions where the system is in a Mott-insulating state as indicated by the thick horizontal lines.}
\label{fig:Hpd_phi}
\end{figure}

\begin{figure*}[tb]
\centering
\includegraphics[width=0.9\linewidth]{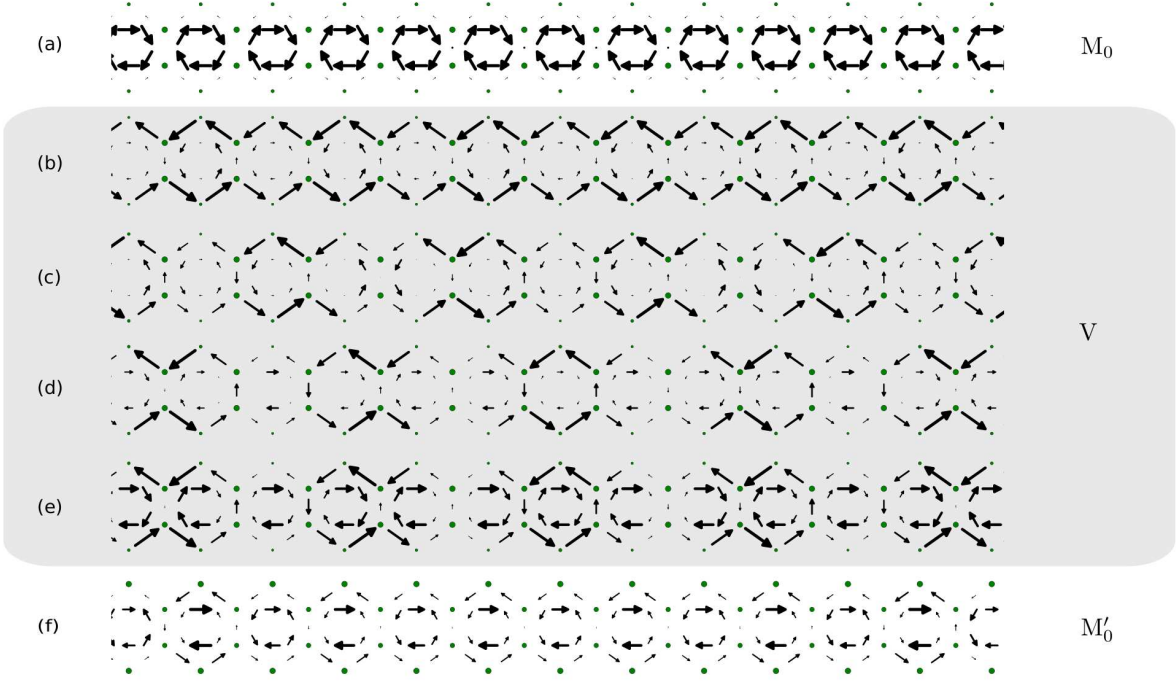}
\caption{(Color online) Current and density configurations of the hardcore boson ground state for different values of $\phi_H$ ($J_H=J$, $\rho=1/4$).
(a) $\phi_H=0.4\pi$ (\Mz phase), (b-e) $\phi_H=0.5\pi$, $0.6\pi$, $0.7\pi$, $0.8\pi$ (V phase) and (f) $\phi_H=0.95\pi$ (\Mzp phase).}
\label{fig:Hpd_phi_patterns}
\end{figure*}

\subsection{Comparison to DMRG results}
Although the Bogoliubov approach is a crude simplification,  we nevertheless obtain a decent qualitative agreement for the phase boundaries of the \VL phase with our numerical DMRG results. We employ a cutoff for the  occupation of bosons per site of typically $n_{\rm max}=4$ bosons for $U\gtrsim J$ and fillings $\rho<1$. By comparison with larger and smaller cutoffs we have ensured the independence of the numerical data on the cutoff for the quantities shown in this work.

The lines of instability of the 
weak-coupling Bogoliubov method (see Fig.~\ref{fig:pd_fU}) predict a linear opening of the \VL phase for $J>J_H^c$, which is consistent with the numerical estimates obtained for a finite filling $\rho=0.2$ and interaction strength  $U\sim J$. Again, for the \Mz to \Mzp transition, we do not resolve any intermediate phase in our numerical simulations.

\section{Phase diagram as a function of $\phi_H$}
\label{sec:phiH}

\begin{figure}[tb]
\centering
\includegraphics[width=1.\linewidth]{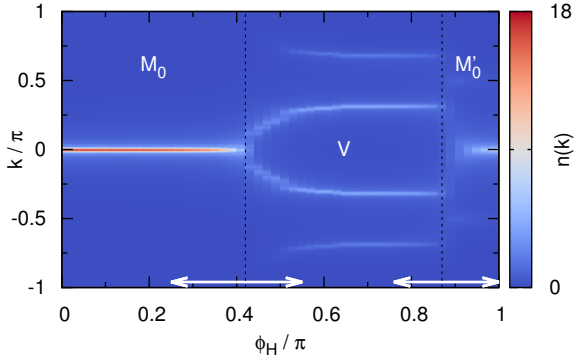}
\caption{(Color online) Quasimomentum distribution function $n(k) $ of hardcore bosons as a function of $\phi_H$  for the parameters of  Fig.~\ref{fig:Hpd_phi} for $\rho=0.25$. The arrows denote the estimated extension of the Mott-insulating regions (see Fig.~\ref{fig:Hpd_phi_cuts}). 
}
\label{fig:Hpd_phi_mom}
\end{figure}

\begin{figure}[tb]
\centering
\includegraphics[width=1.\linewidth]{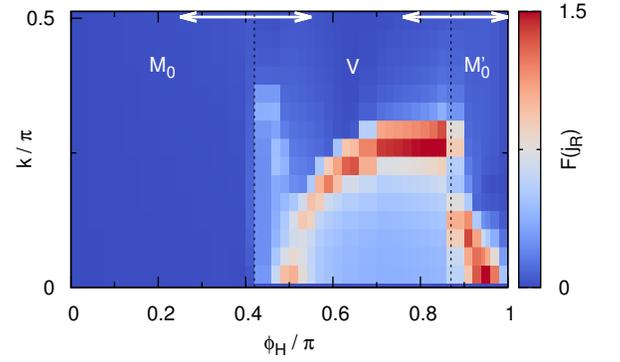}
\caption{(Color online) Fourier transform of the real-space pattern of the rung currents $F(j_R)(k)$ as a function of $\phi_H$ of hardcore bosons for the parameters of  Fig.~\ref{fig:Hpd_phi} for $\rho=0.25$.
}
\label{fig:Hpd_phi_ccf}
\end{figure}

\begin{figure}[tb]
\centering
\includegraphics[width=1.\linewidth]{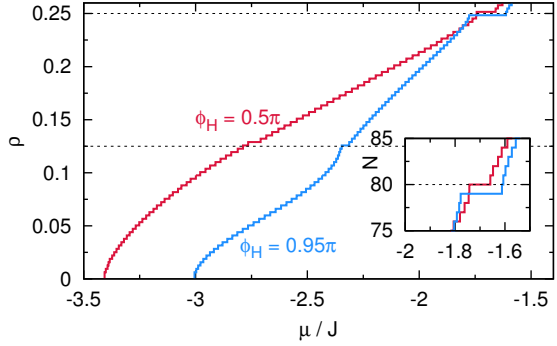}
\caption{(Color online) Examples for the equation of state $\rho=\rho(\mu)$ for two cuts through the phase diagram Fig.~\ref{fig:Hpd_phi} for $\phi_H=0.5\pi$ and $\phi_H=0.95\pi$ ($J_H=J$, $L=159$ rungs). As discussed in the text, due to the open boundary condition in the finite-system size simulations,  the commensurate $\rho=1/4$ MI plateau is found for $N=80$ ($79$) particles for $\phi_H\lesssim 0.9\pi$ ($\phi_H\gtrsim 0.9\pi$). The $\rho(\mu)$ curve for $\phi_H=0.95\pi$ exhibits a small additional plateau close to filling $\rho=1/8$. The inset shows a zoom into the region around $\rho=0.25$ (here, the y-axis shows the total particle number $N$).}
\label{fig:Hpd_phi_mu_rho}
\end{figure}

\begin{figure}[tb]
\centering
\includegraphics[width=1.\linewidth]{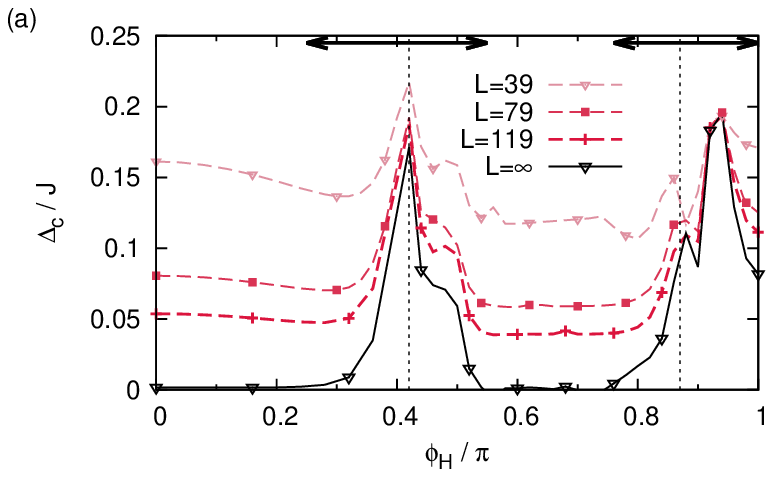}
\includegraphics[width=1.\linewidth]{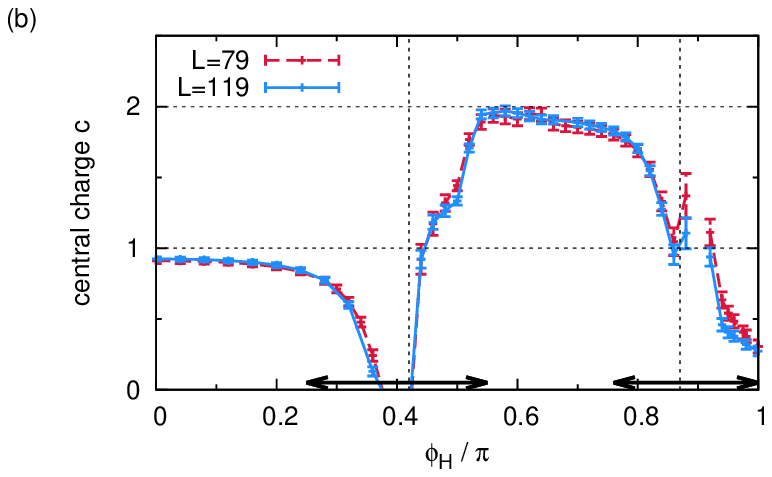}
\includegraphics[width=1.\linewidth]{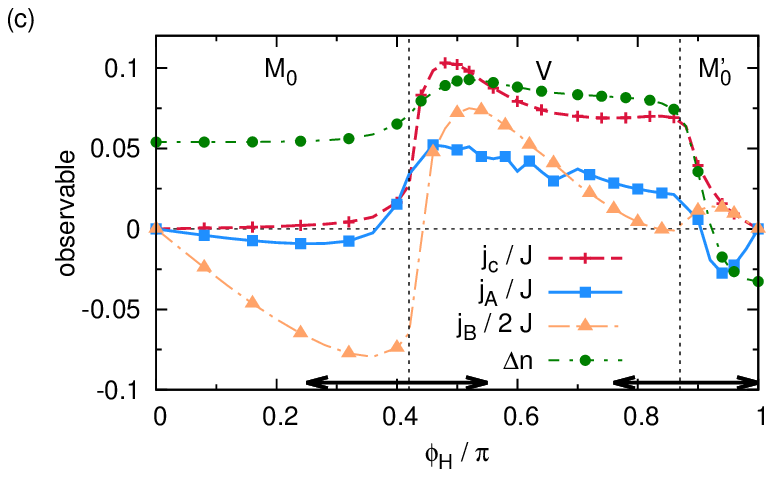}
\caption{(Color online) Cut through the phase diagram of hardcore bosons Fig.~\ref{fig:Hpd_phi} for $\rho=0.25$. 
(a) Charge gap $\Delta_c$ for different system sizes as well as an extrapolation to the thermodynamic limit $L\to\infty$ (using a cubic polynomial in $1/L$).
(b) Central charge $c$ estimated from a fit to Eq.~\eqref{eq:CCformula}.  
(c) Average currents
$\jc$, $j_A$, $j_B/2$, and the density imbalance $\Delta n$.
The arrows (a)-(c) indicate the  extension of the Mott-insulating regions, estimated from a threshold on the extrapolation of the charge gap $\Delta_c / J<10^{-4}$. 
The displayed quantities correspond to fillings of $N=(L+1)/2$ particles for $\phi_H<0.9\pi$ and $N=(L-1)/2$ for $\phi_H>0.9\pi$ (corresponding to the largest finite-size charge gap).
}
\label{fig:Hpd_phi_cuts}
\end{figure}

\begin{figure}[tb]
\centering
\includegraphics[width=1.\linewidth]{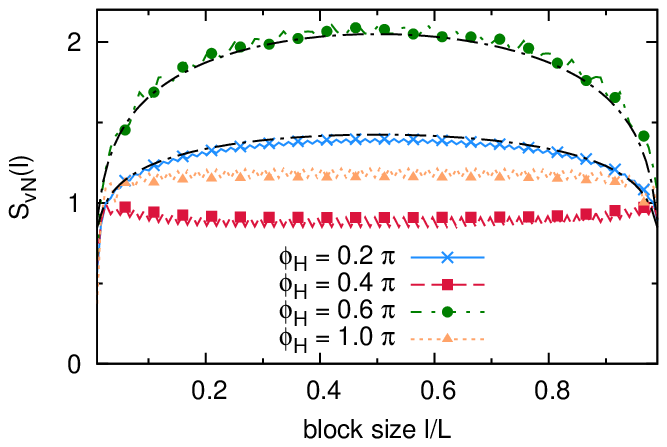}
\caption{(Color online) Examples of the entanglement entropy $S_{\rm vN}$ for various values of $\phi_H$ and hardcore bosons ($J=J_H$, $\rho=0.25$) 
corresponding to the  \Mz-MI ($\phi_H=0.2\pi$),  \Mzp-MI ($\phi_H=0.4\pi$),  \Mz-SF ($\phi_H=0.6\pi$) and V-SF ($\phi_H=\pi$) phases. The black dash-dotted lines correspond to a fit to Eq.~\eqref{eq:CCformula}.
}
\label{fig:Hpd_phi_ee}
\end{figure}

In the range of parameters of Fig.~\ref{fig:pdH2}, we did not find two-component vortex-fluid (V) phases for interacting bosons. However, for different parameters, where a lowest-band minimum exhibits a degeneracy (see Fig.~\ref{fig:sf_disp}~(d)), we observe a bosonic vortex-fluid phase. In Fig.~\ref{fig:Hpd_phi}, we show the ground-state phase diagram for hardcore bosons as a function of the phase $\phi_H$ and the density $\rho$ for $J_H=J$. For $0.381 \lesssim \phi_H/\pi \lesssim 0.85$ the minimum is twofold degenerate (compare Fig.~\ref{fig:sf_disp}(d)). Interestingly, almost independently of the filling $\rho$, a two-component vortex-fluid phase emerges on top of a superfluid background. 
Exemplary current and density configurations for a cut through the phase diagram Fig.~\ref{fig:Hpd_phi} at quarter filling $\rho=1/4$ are shown in Fig.~\ref{fig:Hpd_phi_patterns}.
For large values $\phi_H\sim 0.9\pi$, the \VL phase can again be found.

Due to the approximate independence of the boundary of the vortex-fluid phase from the density, it is difficult to use features of the $\rho(\mu)$ curves (such as shown in Ref.~\cite{Piraud2015}) to extract the position of the phase transition.
The boundary of the vortex-fluid phase, however, can also be extracted  from a calculation of the central charge~\cite{Piraud2015}. This works best for incommensurate fillings -  for the case of commensurate fillings, this becomes more involved as we will discuss in the following.

Experimentally, the vortex-fluid phase may be clearly distinguished from other phases in measurements of the quasimomentum distribution function, in which a multi-peak structure at $\pm Q$ with (in general) $Q\neq 0,\pi/2$ can be observed. We show the corresponding quasimomentum distribution in Fig.~\ref{fig:Hpd_phi_mom}.

In Fig.~\ref{fig:Hpd_phi_ccf} we also plot the Fourier transform $F(j_R)$ of the real-space patterns of the rung currents $\langle \mathcal{J}(A_{1,\ell}\to A_{2,\ell}) \rangle(\ell)$. Its distinct peak position may be interpreted as a measure of the vortex-density~\cite{Piraud2015} in the system. As previously shown in Ref.~\cite{Greschner2016} both $F(j_R)$ and the momentum distribution show a similar behavior. 

The peak position of $F(j_R)$ in Fig.~\ref{fig:Hpd_phi_ccf} exhibits a sharp jump for $\phi_H\approx 0.86 \pi$, which we here identify as the V to \Mzp transition point.
Close to this V to \Mzp boundary the quasimomentum distribution of Fig.~\ref{fig:Hpd_phi_mom} becomes blurred. Interestingly, in this part of the \Mzp region,  $F(j_R)$ also exhibits a distinct peak at $k>0$, i.e., finite (boundary driven) oscillations of the rung-currents can still be found.
Similar incommensurate Meissner-like-phases have been discussed in Ref.~\cite{Greschner2016} and have been connected to a certain class of Laughlin-precursor states~\cite{Petrescu2017,Strinati2016} for the case of two-leg flux ladders. Indeed, the presence of such further intermediate phases close to the V to \Mzp boundary in this model should be examined in future studies more in detail. 

For commensurate fillings, $\rho=1/8$ and $\rho=1/4$, we observe the opening of a charge gap for certain values of $\phi_H$. This can be the best seen in the $\rho(\mu)$ curves displayed in Fig.~\ref{fig:Hpd_phi_mu_rho} for different values of $\phi_H$, where small horizontal plateaus at fillings $\rho=1/4$ and $\rho=1/8$ indicate the MI-regions. 

In Fig.~\ref{fig:Hpd_phi_cuts}~(a) we show the extracted charge gap 
\begin{align}
\Delta_c = \frac{E_0(N-1)-2 E_0(N) + E_0(N+1)}{2} \,,
\end{align}
extrapolated to the thermodynamic limit $L\to\infty$. Due to the effects of the open boundary conditions the 
particle density corresponding to the MI-plateau is slightly offset from commensurability, depending on the choice of parameters. We display the data for $N=(L\pm 1)/2$ particles, which corresponds to the largest finite-size value of $\Delta_c$.

Therefore, again, we observe the \Mz and \Mzp states and here also the vortex-fluid phases in both the SF and the MI background.
 For the case of a V - MI phase, we expect the presence of a gapless neutral excitation and, hence, a central charge $c=1$. In Fig.~\ref{fig:Hpd_phi_cuts}(c) we show the extracted central charge from fits to the entanglement entropy. Examples for the entanglement entropy and its dependence on block size  are shown in Fig.~\ref{fig:Hpd_phi_ee}. The results are consistent with $c=0$ in the \Mz- MI and \Mzp- MI phases, $c=1$ in the V- and \Mz-SF phases and $c=2$ in the V-SF phase.

The horizontal arrows in Figs.~\ref{fig:Hpd_phi_mom}, \ref{fig:Hpd_phi_ccf} and \ref{fig:Hpd_phi_cuts} show the estimated extension of the MI phases. Due to the Berezinskii-Kosterlitz-Thouless nature of the Mott-insulator to superfluid phase transitions we only give an approximate extension based on the extrapolation of the charge gap and the calculation of the central charge.

In Fig.~\ref{fig:Hpd_phi_cuts}~(c), we plot the behavior of chiral currents and the density imbalance for a cut through the phase diagram Fig.~\ref{fig:Hpd_phi} at the commensurate filling $\rho=1/4$. Consistent with our previous observations,  we also find the characteristic features of the Meissner phases: For the \Mz phase, we find $j_c$ and $j_A\approx 0$ and $j_B<0$ as well as $\Delta n>0$, while for the \Mzp phase we mainly observe opposite signs, $j_B>0$ and $\Delta n<0$.

\section{Summary}
\label{sec:sum}

In summary, we have systematically	 studied the ground-state phase diagram of interacting bosons (and free fermions) for the Haldane model on a minimal realization of a two-leg ladder. Our main result is the emergence of an exotic type of a vortex-lattice like phase for interacting bosons even for hardcore interactions. The \VL phase exhibits a finite rung-current order parameter as well as a finite charge-density wave ordering. Since it emerges both at commensurate fillings with a charge gap but also at {a broad range of} incommensurate fillings on a superfluid gapless background, in the latter case, the \VL phase can be understood as another example of a lattice supersolid, i.e., a liquid with charge-density ordering.
 
We conclude by pointing future research directions related to this model. In particular, the presence of analogs of Laughlin-precursor states discussed in Refs.~\cite{Petrescu2017,Strinati2016} may be examined in the region between the vortex-fluid and Meissner phases in the future. 
Further possible  extensions include   the analysis of quantum phases in extended lattice geometries such as three-leg ladders and  simplified (i.e., no next-to-nearest-neighbor tunneling)  brick-wall ladders with a flux that are the thin-torus limit of the hexagonal lattice along the zigzag cut.

\begin{acknowledgments}
We are grateful to L.~Santos and T.~Vekua for useful discussions and we are indebted to G. Roux for his helpful comments on
a previous version of the manuscript. S.G. acknowledges support from the German Research Foundation DFG (project no. SA 1031/10-1). 
F.H.-M. acknowledges support from the DFG (Research Unit FOR 2414) via grant no. HE 5242/4-1.
Simulations were carried out on the cluster system at the Leibniz University of Hannover, Germany.
The work of F.H.-M. was performed in part 
at the Aspen Center for Physics 
which is supported by National Science Foundation
Grant No. PHYS-1607611. The hospitality of the Aspen Center for Physics is gratefully acknowledged.
\end{acknowledgments}

\bibliography{references}

\end{document}